# Comparative measurements of ITER Nb$_3$Sn strands between two laboratories

J. Lu, S. Hill, D. McGuire and K. Dellinger (National High Magnetic Field Laboratory, USA),

A. Nijhuis, W.A.J. Wessel and H.J.G. Krooshoop (University of Twente, Netherlands),

K. Chan and N.N. Martovestky (Oak Ridge National Laboratory, USA)

**Abstract**

ITER Nb$_3$Sn strand quality verification tests require large quantities of precise measurements. Therefore regular cross-checking between testing laboratories is critically important. In this paper, we present results from a cross-checking test of 140 samples between the National High Magnetic Field Laboratory, USA and the University of Twente, the Netherlands. The tests comprise measurements at 4.2 K on critical current, RRR and hysteresis loss, while at room temperature the chromium layer thickness, Cu/nonCu ratio, filament twist pitch, and diameter were determined. Our results show very good agreement between the two labs. The reasons for small random discrepancies are discussed.

1.  **Introduction**

Large superconducting magnets such as those for the ITER International Organization require commercial production of a large quantity of Nb$_3$Sn strands [1]. It demands accurate and cost-effective quality assurance measurements to assure reliable operation of the magnets. While testing superconducting strands is very important for large superconducting magnet projects in general [2-5], it is particularly crucial for ITER Nb$_3$Sn strands, which are supplied by multiple strand manufacturers. Therefore a verification test program is implemented involving a number of labs worldwide [6-17].

Historically, Nb$_3$Sn wire manufacturers and various research laboratories use slightly different testing protocols. Therefore the ITER international organization organized benchmarking and annual cross-checking tests to be performed at each testing lab, which carries out heat treatment and quality property verification tests of a few samples [6, 17]. These tests are typically performed on samples cut from a single piece length of $Nb_3Sn$ wire assuming sufficient quality and homogeneity of the properties along the wire length. Due to the small number of test samples for each participating lab, analysis of the statistical difference between labs is difficult. This paper presents the results of a special set of cross-checking measurements of large number of production samples which is considerably more than number of ITER routine cross-checking samples. In this experiment, samples from 140 billets made by one manufacturer are prepared, heat treated, and tested by both the National High Magnetic Field Laboratory, USA (NHMFL) and the University of Twente (UT) in a production testing mode. We compare the test results of critical current ($I_c$), residual resistance ratio (RRR), hysteresis loss ($Q_{hyst}$), diameter, chromium coating thickness, twist pitch and copper to non-copper ratio between two laboratories. The relatively large number of tests allows us to perform statistical analysis and to identify the significance of the difference or lack thereof between the two sets of data.

2. **Experimental methods**

$Nb_3Sn$ strands used in this experiment are designed for the ITER Toroidal Field (TF) coils and manufactured by the internal-tin process by Luvata (Waterbury, CT, USA). A cross-section of an unreacted strand is shown in figure 1. A 20 m long wire is cut from each of the 140 billets for testing. NHMFL and UT each performed independent heat treatment, liquid helium temperature testing and room temperature testing.

Reaction heat treatment at both labs were performed in argon gas following ITER heat treatment schedule B,

    210 C for 50 h

    340 C for 25 h

450 C for 25 h

575 C for 100 h

650 C for 100 h

cool to 500 C + furnace cool

The ramp rate = 5 C/h for all ramps.

At NHMFL, heat treatment of in total 140 samples was done in consecutive 6 batches. While the heat treatment at UT was completed in one batch. Some details of the measurement techniques adopted at the NHMFL are given in [7] and of the UT test techniques in [11, 12]. For the ease of comparison, a brief description of test methods used by each lab is listed in Table I, which also contains the number of samples for each test. The RRR of ITER $Nb_3Sn$ strand is defined as its resistance ratio between 273 and 20 K. Both $I_c$ and $n$ were measured at 11.5, 12.0 and 12.5 T, but only 12.0 T data are presented in this paper. $I_c$ self-field corrections were not applied.

3. Results and Discussions

A comparison of $I_c$ measured by NHMFL and UT is shown in figure 2(a). It is evident that these two sets of data are in good agreement with one another. The $I_c$ difference between UT and NHFML, $I_c$_UT – $I_c$_NHMFL ($\Delta I_c$), is also plotted for each sample in this figure. The $\Delta I_c$ varies randomly around zero, and the variation of $\Delta I_c$ from billet to billet is somewhat smaller than the variation of $I_c$. Figure 2(b) shows $I_c$ measured by UT against $I_c$ measured by NHFML. There is no significant systematic difference between UT and NHMFL $I_c$ data. It is noted, however, that a few $\Delta I_c$ values are as large as 30 A. Since the uncertainty in $I_c$ measurement is only about 1 A [18], and there is no evidence of discrepancy in heat treatment which would cause a systematic difference between UT and NHMFL $I_c$, we suspect that the large $\Delta I_c$ is due to the appreciable property variation within the 20 m sampling length for each billet. This hypothesis is supported by the fact that $I_c$ variation within a billet can be as much as 30 A for wires made by multiple manufacturers including Luvata [19].

Similar plots of *n* value, RRR and $Q_{hyst}$ comparisons are shown in figure 3, 4 and 5 respectively. Similar to the case with $I_c$, the agreement between UT and NHMFL is good. The relatively larger variations in $\Delta n$, $\Delta RRR$ and $\Delta Q_{hyst}$ cannot be explained by the uncertainty of measurement techniques which is about 1, 1 and 3 kJ/m$^3$ for $\Delta n$, $\Delta RRR$ and $\Delta Q_{hyst}$ respectively [20]. Again they might be attributed to the non-uniformity of properties within 20 m of sample for each billet, and are supported by observed variations in $\Delta n$, $\Delta RRR$ and $\Delta Q_{hyst}$ within one billet [19]. Finally, a summary of results and statistics for all tests including room temperature test is shown in Table II. The difference in mean values between NHMFL and UT data is small as compared with random data scattering quantified by standard deviation, except for the diameter measurement which requires further investigation.

4. Conclusion

NHMFL and UT conducted cross-checking measurements of 140 internal-tin Nb$_3$Sn wires. Samples were heat treated and tested in each laboratory separately. Comparison of $I_c$, *n*, RRR, and $Q_{hyst}$ are presented. The results from the two labs are in good agreement. The random difference is attributed to the variation along the sampling wire length. This cross-checking of large number of samples provides a set of interesting and reassuring data which confirms statistically significant agreement between two labs.

5. Acknowledgement

The authors from NHMFL thank financial supports in part by the U.S. Department of Energy vie US-ITER under subcontract 4000110684, by the National Science Foundation under Grant DMR-0084173, and by the State of Florida.

The authors from UT wish to acknowledge the support from the ITER International Organization and F4E Barcelona. Disclaimer: The views and opinions expressed herein do not necessarily reflect those of F4E or the ITER International Organization.

The authors from US-ITER would like to thank the U.S. Department of Energy by Lawrence Livermore National Laboratory under Contract DE-AC52-07NA27344, by UT-Battelle, LLC, under contract DE-AC05-00OR22725 with the U.S. Department of Energy, and by the U.S. Department of Energy, Office of Science, Office of Fusion Energy Sciences.

**Table and figure captions**

Table I Test methods used by NHFML and UT

Table II Summary of comparative test results

Figure 1 Cross-section of an unreacted $Nb_3Sn$ strand used in this experiment.

Figure 2 $I_c$ measured at 4.2 K 12 T, comparison between NHMFL and UT for 140 samples. (a) the left vertical axis is for $I_c$ measured by NHMFL and UT, the right vertical axis is for the $I_c$ difference between UT and NHMFL. (b) UT $I_c$ versus NHMFL $I_c$, the diagonal line indicates an ideal correlation between the two sets of data.

Figure 3 $n$ value measured at 4.2 K 12 T, comparison between NHMFL and UT for 140 samples. (a) the left vertical axis is for $n$ measured by NHMFL and UT, the right vertical axis is for the $n$ difference between UT and NHMFL. (b) UT $n$ versus NHMFL $n$, the diagonal line indicates an ideal correlation between the two sets of data.

Figure 4 RRR comparison between NHMFL and UT for 80 samples. (a) the left vertical axis is for RRR measured by NHMFL and UT, the right vertical axis is for the RRR difference between UT and NHMFL. (b) UT RRR versus NHMFL RRR, the diagonal line indicates an ideal correlation between the two sets of data.

Figure 5 $Q_{hyst}$ comparison between NHMFL and UT for 40 samples. (a) the left vertical axis is for $Q_{hyst}$ measured by NHMFL and UT, the right vertical axis is for the $Q_{hyst}$ difference between UT and NHMFL. (b) UT $Q_{hyst}$ versus NHMFL $Q_{hyst}$, the diagonal line indicates an ideal correlation between the two sets of data.

Table I. Test methods used by NHMFL and UT.

| | # of tests | NHMFL | UT |
|---|---|---|---|
| Heat treatment | 140 | In argon, ITER schedule B | In vacuum, ITER schedule B |
| $I_c$ | 140 | ITER barrel, no handling after HT, 25 and 50 cm taps | ITER barrel, no handling after HT, 25 and 50 cm taps |
| RRR | 80 | 150 mm straight sample, natural warming to 20 K | 100 mm straight sample, natural warming to 20 K |
| $Q_{hyst}$ | 40 | VSM, 7 turn 4 mm diameter coil | Magnetometer, 14 turn, 40 mm diameter coil, 2 m wire length |
| Cr thickness | 80 | Light microscopy | Etching, weigh |
| Cu/non Cu | 80 | Light microscopy | Etching, weigh |
| Twist Pitch | 140 | Etch, incline angle | Etch, incline angle |
| Diameter | 80 | Digital micrometer | Digital micrometer |

Table II. Summary of test results.

| Test | NHMFL | | UT | | UT - NHMFL | |
|---|---|---|---|---|---|---|
| | Mean | Standard deviation | Mean | Standard deviation | Mean | Standard deviation |
| $I_c$ (12 T) (A) | 247.4 | 13.2 | 249.7 | 13.6 | 2.3 | 6.9 |
| $n$ | 20.8 | 2.1 | 20.8 | 2.4 | 0.1 | 1.2 |
| RRR | 154.4 | 13 | 154.8 | 11.3 | 0.4 | 9.0 |
| $Q_{hyst}$ (kJ/m$^3$) | 350.0 | 50.8 | 341.5 | 39.1 | -8.4 | 43.6 |
| Cr thickness (μm) | 1.3 | 0.3 | 1.3 | 0.3 | 0.0 | 0.2 |
| Cu/non Cu | 0.916 | 0.012 | 0.902 | 0.014 | -0.013 | 0.016 |
| Twist Pitch (mm) | 15.4 | 0.8 | 16.8 | 1.2 | 1.5 | 1.5 |
| Diameter(mm) | 0.819 | 0.001 | 0.821 | 0.001 | 0.003 | 0.001 |

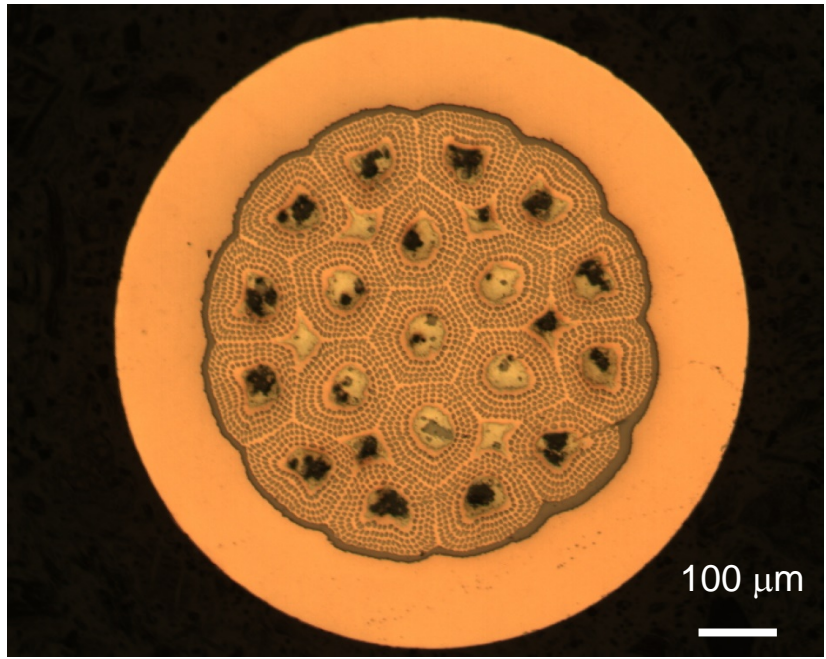

Lu, et al. Fig. 1

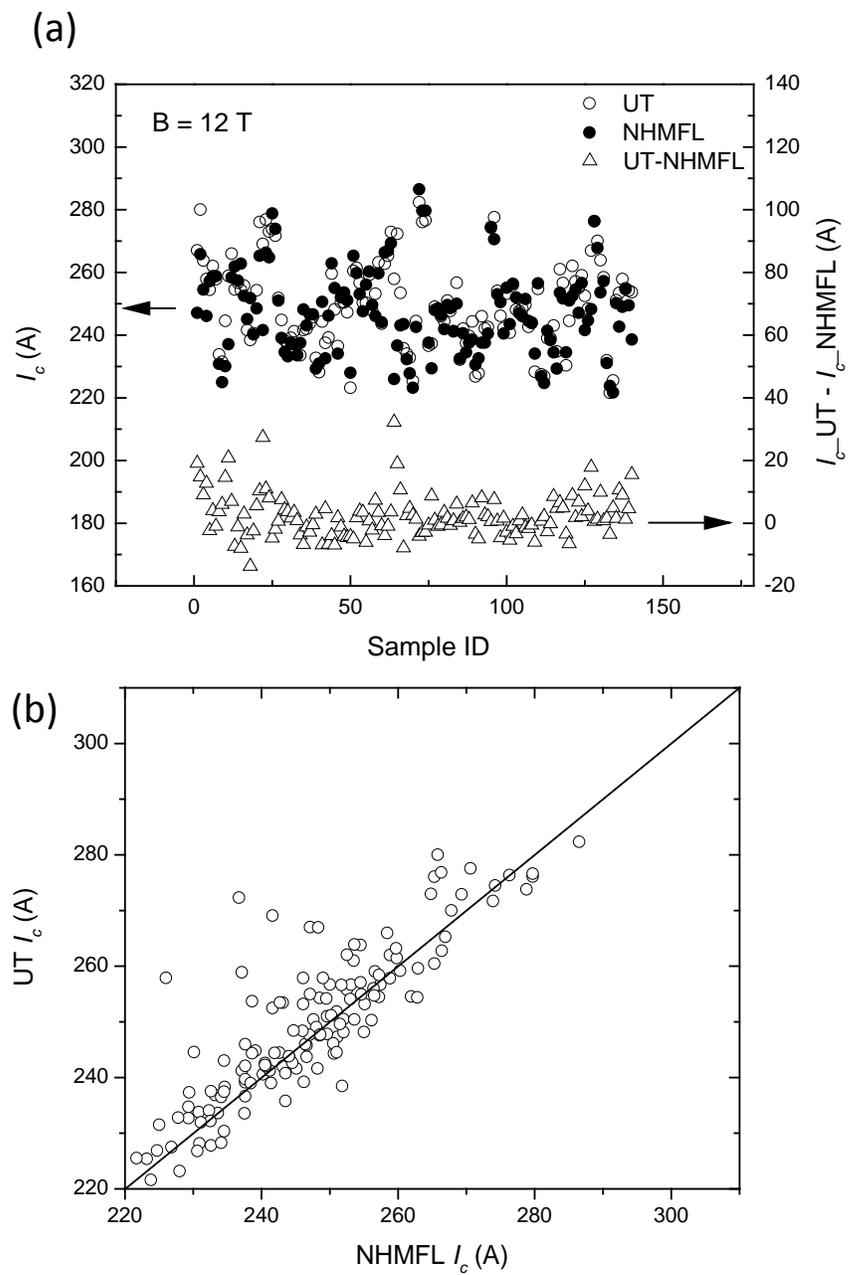

Lu, et al. Fig. 2

(a)

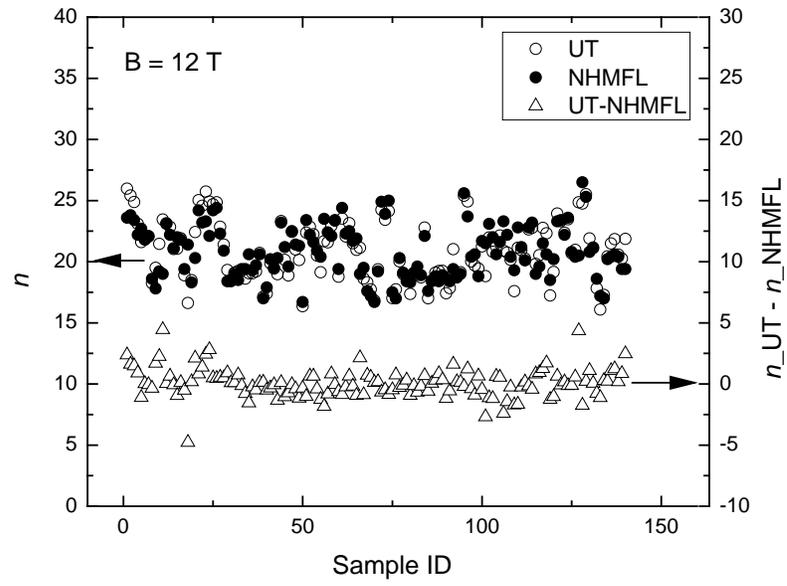

(b)

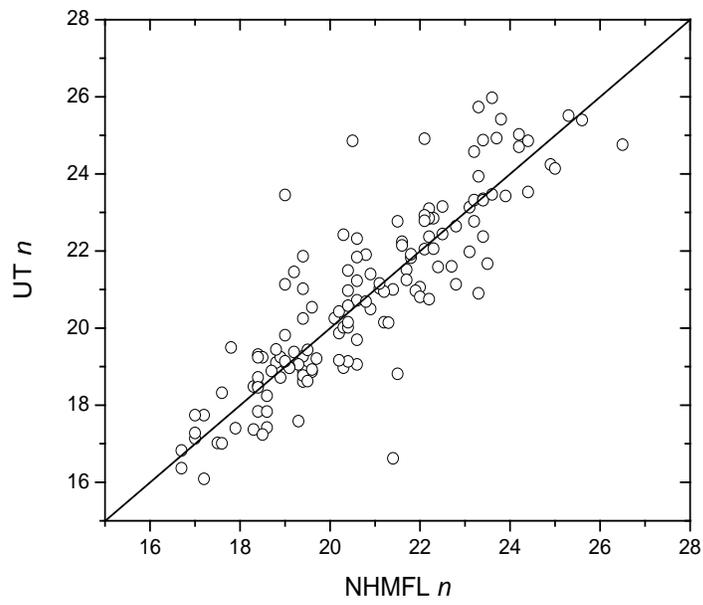

Lu, et al. Fig. 3

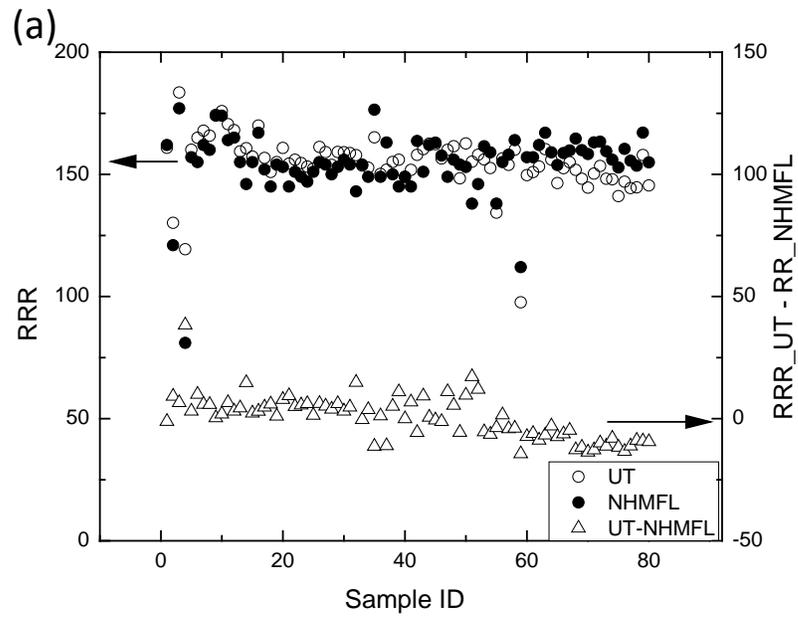

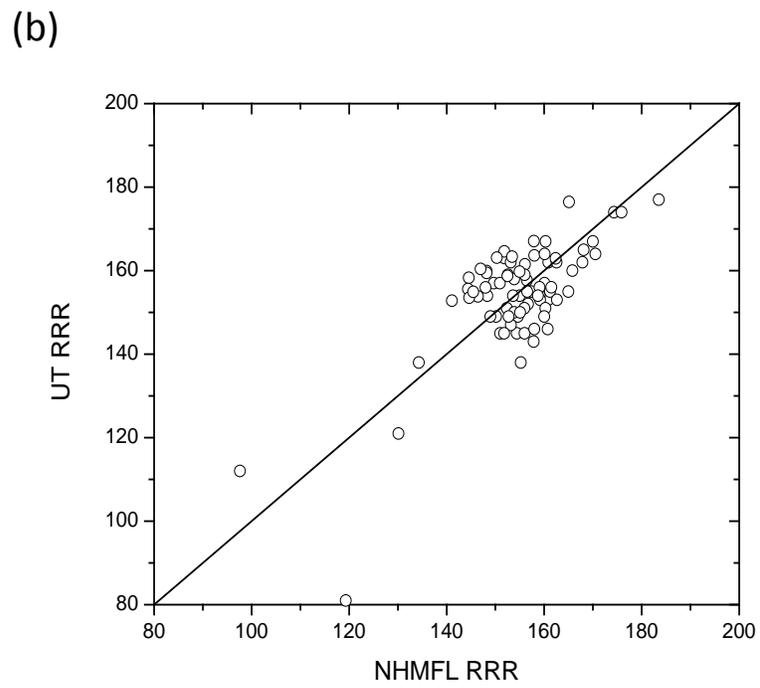

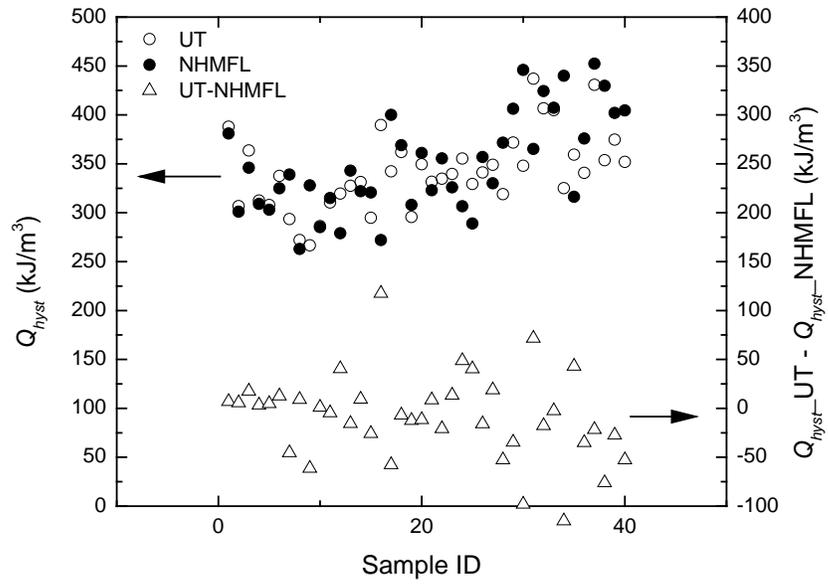

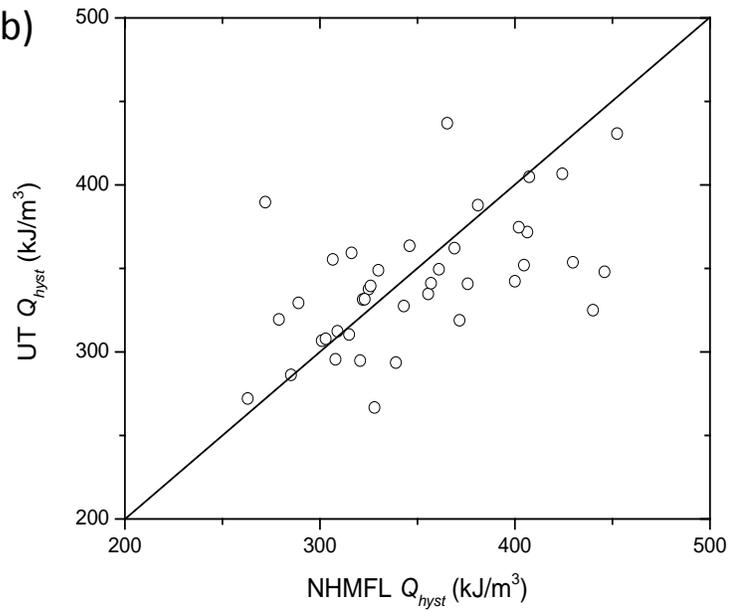

Lu, et al. Fig. 5